\documentclass[reprint,amsmath,amssymb,aps,]{revtex4-2}

\usepackage{graphicx}
\usepackage{dcolumn}
\usepackage{bm}
\usepackage{color}
\usepackage{hyperref}
\usepackage{soul}

\begin{document}


\title{Flapping, swirling and flipping: Non-linear dynamics of pre-stressed active filaments} 

\author{Soheil Fatehiboroujeni}
 \email{sfatehiboroujeni@ucmerced.edu}
 \affiliation{Department of Mechanical Engineering, University of California, Merced}
  \author{Arvind Gopinath}
 \email{agopinath@ucmerced.edu}
\affiliation{Department of Bioengineering, University of California Merced, Merced CA. \\
Health Sciences Research Institute, University of California, Merced}
\author{Sachin Goyal}
 \affiliation{Department of Mechanical Engineering, University of California Merced, Merced CA. \\
 Health Sciences Research Institute, University of California, Merced}

\date{\today}
            
 \begin{abstract}
Initially straight slender elastic rods with geometrically constrained ends buckle and form stable two-dimensional shapes when compressed by bringing the ends together. It is also known that beyond a critical value of the pre-stress, clamped rods transition to bent, twisted three-dimensional equilibrium shapes. Recently, we showed that pre-stressed planar shapes when immersed in a dissipative fluid and animated by nonconservative follower forces exhibit stable large-amplitude flapping oscillations. Here, we use time-stepper methods to analyze the three-dimensional instabilities and dynamics of pre-stressed planar and non-planar filament configurations when subject to active follower forces and dissipative fluid drag. 
First, we find that type of boundary constraint determines the nature of the non-linear patterns following instability.
When the filament is clamped at one end and pinned at the other with follower forces directed towards the clamped end, we observe only stable planar (flapping) oscillations termed flapping result. 
When both ends are clamped however, we observe a secondary instability wherein planar oscillations are destabilized by off-planar perturbations and result in fully three-dimensional swirling patterns characterized by two distinct time-scales. The first time scale characterizes continuous and unidirectional swirling rotation around the end-to-end axis. The second time scale captures the rate at which the direction of swirling reverses or flips. The overall time over which the direction of swirling flips is very short compared to the long times over which the filament swirls in the same direction. Computations indicate that the reversal of swirling oscillations resembles relaxation oscillations with each cycle initiated by a sudden jump in torsional deformation and then followed by a period of gradual decrease in net torsion until the next cycle of variations. Our work reveals the rich tapestry of spatiotemporal patterns when weakly inertial strongly damped rods are deformed by non-conservative active forces. Practically, our results suggest avenues by which pre-stress, elasticity and activity may be used to design synthetic fluidic elements to pump or mix fluid at macroscopic length scales. The connection to relaxation oscillations motivates future theoretical work that may yield  principles to realize and tune these designs. 

\end{abstract}

                             
\maketitle

\section{\label{sec:int}Introduction}

The deformation of slender rods and filaments has been the subject of intense and productive enquiry since 1757 when Euler investigated the instability of loaded slender elastic columns and beams \cite{doi:10.1086/346767}
and derived estimates of the critical axial load that results in the instability and large amplitude deformation now known as Euler buckling. The estimate of the critical load for buckling was further refined using higher order theory by Lagrange in 1770  \cite{todhunter1893history,timoshenko1983history}.  Paraphrasing Truesdell \cite{euler1960rational}, these seminal studies and the theoretical foundation of most of what followed were grounded in two main ideas - the first of Hooke (1678) proposed
that the displacement of a elastic body was in proportion to the load causing the
displacement, and second Bernoulli's hypothesis (1705) that the curvature at any point in a bent rod was in
proportion to the resisting moment developed locally in the rod. 
Since then, Euler buckling has played a central role in understanding the mechanics and dynamics of slender structures across disciplines in engineering
\cite{timoshenko2009theory,Nizette1999, Antman2005,1531-3492_2003_4_505,ogden1997non,biot:hal-01352219,10.1083/jcb.120.4.923,Weber10703, Gopinath2011}, botany \cite{niklas1992plant}, and biophysics \cite{cite-key-howard2, cite-key-gibbons, cite-key-howard, Vaziri2008, cite-key-Ainsworth, Bayly2016,BAYLY20141756, cite-key-gibbons, Robisonaaf0659}.

Over the last many decades (running to centuries), theories building on these fundamental concepts have been proposed to analyze instabilities in structures such as cantilevered beams or rods under two types of external forces or loads. 
The first type comprises of
conservative forces derivable from a potential such as gravity \cite{cite-key-Keller,10.2307/24901814}. For such forces, stability properties can be determined by examining an energy functional that encapsulates the incremental changes in the energy of the system and work done at boundaries during the deformation. Mathematically, this implies that static shapes are derived by minimizing a suitably derived Hamiltonian founded on a self-adjoint formulation - an approach that has been explored and utilized intensively (see \cite{cite-key-Keller,10.2307/24901814}, and discussion in  
\cite{Elishakoff2005}).

More recently, attention has turned to the 
study of elastic structures and systems under the action of active {\em non-conservative forces} that cannot be derived from a potential. This includes follower forces - that is forces always aligned along the centerline of the deforming filament and therefore move with the filament as it deforms. The primary motivation for studying these  came initially from problems in aeroelasticity \cite{10.1115/1.4042324} and flow-induced energy harvesting \cite{PIGOLOTTI2017116}. 

Recent focus has however been on problems involving follower forces in bioinspired systems comprised of soft filaments immersed in a fluidic medium. Motivated by the manner in which molecular motors in organelles such as eukaryotic cilia and flagella drive oscillatory motions and wave-like beating with distinct frequencies and wavelengths \cite{doi:10.1146/annurev.fl.09.010177.002011,doi:10.1098/rsif.2018.0594,zhang:04a,doi:10.1002/cm.970140305,doi:10.1002/cm.20313,doi:10.1111/j.1469-185X.2009.00110.x,cite-key11,Lindemann519}, significant work has gone into developing synthetic mimics that generate similar functionality such as locomotion, mixing and mechanosensing \cite{bookBray, cite-key, Malone13325, Gopinath2011}. 
Abstracting the key components of the immensely complicated biological systems,  these analogues exploit combine fluid dissipation, and elasticity with active forces originating from magnetic, electric or chemical fields to induce deformation, buckling and motion \cite{fluidmanip,Dreyfus2005,Nishiguchi_2018,PATTESON201686,Sasaki2014}. At larger,  macroscopic length scales, a recent experimental setup \cite{BIGONI201899} is also designed and validated to introduce follower tangential forces at the end of an elastic rod by exploiting Coulomb friction. We note that the ambient fluid medium exerts a drag on these soft filaments as they deform. Notably, a component of this drag is tangential and thus constitutes a passive con-conservative follower force. 

In contrast to filaments subject to conservative forces, stability  characteristics of structures animated by non-conservative follower forces cannot be examined by energy based arguments. This is due to the non-self-adjoint nature of the equations and boundary conditions \cite{cite-key-Keller,10.2307/24901814,Leipholz01,doi:10.1139/l90-034,cite-keyBolotin,Pfluger01,Ziegler1977, 10.2307/2416522,Bolotin99,Elishakoff2005}. Instead time dependent evolution equations (dynamical equations) derived using variants of the Kirchhoff-Love theory  are appropriate in this situation and have been used to analyze the onset of instabilities and non-linear patterns in these systems \cite{DeCanio20170491,doi:10.1098/rsif.2018.0594,PMID:24352670, Arvind20a,Sangani2020.03.10.986596}. 

An significant gap in the literature and in the understanding of the mechanics and dynamics of active filaments however exists.  Most current theories relate to filaments/rods that are are not pre-stressed (equivalently not pre-strained) and further are only partially constrained; that is, in most cases the base state is an uncompressed rod with a straight shape  \cite{DeCanio20170491,doi:10.1098/rsif.2018.0594,PMID:24352670, Arvind20a,Sangani2020.03.10.986596}. In many important contexts however, rods and filaments that are subject to deforming forces and torques start off from shapes that are neither planar nor stress-free. Such {\em pre-stressed and twisted three-dimensional shapes} abound in nature at all scales; examples include the buckling of growing tendrils \cite{Liu7100,PhysRevLett.80.1564}, the curling following bending of ropes hitting a plane surface \cite{Jawed14663}, torsionally constrained DNA looping mediated by protein binding  \cite{balaeff:06a,goyal:08b}, self-contact driven buckling of DNA  \cite{forth:08a,goyal:08c}, and relaxation of DNA supercoils by topoisomerases \cite{cite-key-Lillian}. It is worth nothing that significant efforts towards accurate modeling of these is directed towards elucidating material constitutive laws \cite{palanth, soheil20a}. 

As a typical example, we note that a pre-stressed filament clamped at both ends - being both pre-stressed and strongly constrained at the boundaries - is expected to have different dynamics than a stress-free cantilever.  
In recent work, \cite{soheil18a} we used a continuum nonlinear rod model and analyzed the emergence of planar instabilities of a pre-stressed (planarly buckled) filament subject to  non-conservative follower forces. 
In this case, even though both ends of the filament were clamped,  
the  initial compression provides  sufficient slack and furnishes  the necessary degree of freedom for stable planar oscillations to emerge. The main findings of this work was that (a) the critical force for the onset of flapping is strongly dependent on the initial slack and hence the magnitude of the pre-stress, and that (b) far from the critical point, flapping frequencies scale with the intensity of follower force through a power law relationship. These studies complement both continuum and discrete analyses of the dynamics of stress-free filaments animated by follower forces 

In this article we build on this initial study. Keeping the densities  of the rod and fluid constant, we focus on the role of elasticity and on activity in selecting the eventual stable state from an initial stationary configuration. We find a variety of possible stable non-linear solutions that involve purely two-dimensional flapping states or fully three dimensional involving swirls and flips. To the best of our knowledge, such a completely non-linear and fully three-dimensional analysis has not been presented elsewhere. 

The organization of this article is as follows. In section $\S$-II, we present the continuum rod model for animated geometrically nonlinear slender elastic rods also subject to fluid drag.  In section $\S$-III, we summarize results for filaments with planar pre-stressed base states, and then we move to 
results for base states that are fully three-dimensional. To isolate the effect of  boundary conditions from the role of pre-stress and activity, we present here  results for two practically important types of boundary constraints - clamped-clamped rods and pinned-clamped rods constraints in planar post-buckling regime. We then turn to rods with clamped-clamped constraints in secondary bifurcation (bent and twisted) regime. We find that non-planar base states if activated by any non-zero follower force give rise to a swirling (purely rotational) motion around the end-to-end axis of the rod. In addition, we find that swirling oscillations undergo a sudden reversal of direction, i.e., a periodic flipping akin to the phenomena of relaxation oscillations seen in diverse fields such as lasers, cellular phenomena  and in electronic circuits \cite{doi:10.1002/9783527617586.ch1}. 
We conclude in Sections $\S$-IV \& V with a summary of results, their significance and suggestions to extend current work and motivate future theoretical work.

\begin{figure*}
\centering
\includegraphics[width=2\columnwidth]{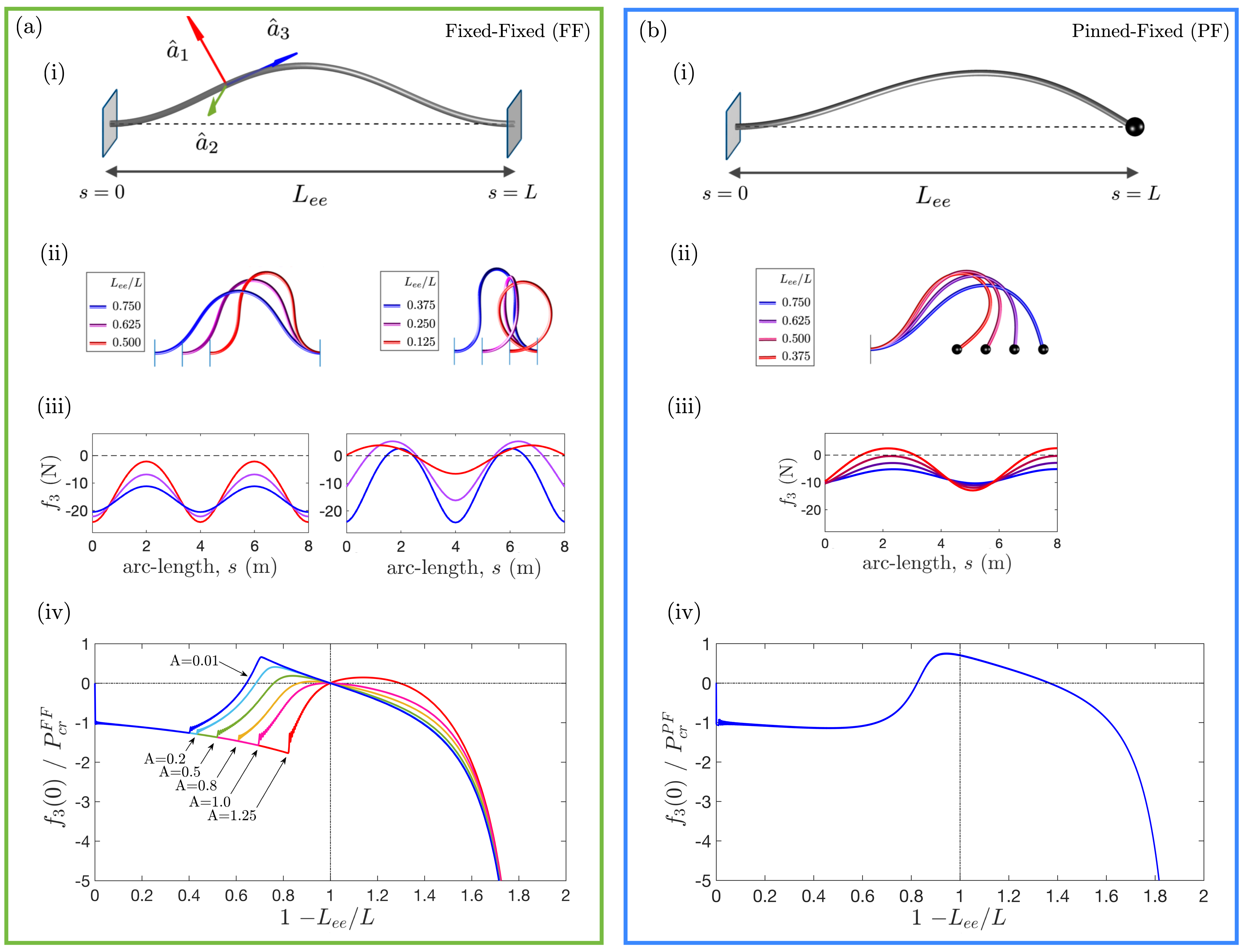}
\caption{We summarize computational results relating to two and three dimensional buckling deformations of a passive elastic slender rod (filament) of length $L$ when pre-stressed by pushing the ends together. The ratio of torsional to bending stiffnesses is $A$. The elastic thin filament we consider has length $L$ and is comprised of a linearly elastic material with Young's modulus $E$. The diameter of the rod $d$ is assumed to be uniform and equal to $\epsilon L$ with $\epsilon \ll 1$ (see Table 1); furthermore we assume the rod to be inextensible and unshearable. This allows us to analyze the spatiotemporal deformations using a  reduced dimensional form of the Kirchhoff equations as in $\S$2. To specify the shape of the filament, we define a reference sphere-fixed coordinate system defined by unit vectors ${\hat{\bf a}}_{1}$, ${\hat{\bf a}}_{2}$ and ${\hat{\bf a}}_{3}$ as shown. Tile (a) summarizes results for a fixed-fixed (FF) boundary condition while tile (b) summarizes 
results for pinned-fixed (PF) boundary conditions.  We compare the response for both types of boundary conditions. 
(a)-(i) is a schematic representations of a pre-stressed, buckled fixed-fixed rod with end-to-end distance $L_{\mathrm{ee}} < L$ while (b)-(i) is the corresponding sketch for the pinned-fixed rod. Arc-length $s$ parametrizes the location of Lagrangian material points in both.
As illustrated in (a)-(ii) and in (b)-(ii), the initially straight filament buckled to a stable, static planar shape when load on the boundary, $f_3(0)$ - the component along ${\hat{\bf a}}_{3}$ - reaches the Euler buckling value. Note that here, this compression or pre-stress is controlled by the value of the slack defined as $1-L_{ee}/L$. The critical values are $P_{\mathrm{cr}}^{\mathrm{FF}}=4\pi EI/L^2$ for fixed-fixed and $P_{\mathrm{cr}}^{\mathrm{PF}}=2.045\pi EI/L^2$ for pinned-fixed conditions. Fixing $A=0.8$, we determined the evolution of static shapes as well as distribution of internal forces in the tangential direction, $f_3$, as a function of $L_{ee}/L$ for both FF and PF cases. These are shown 
in rows (a,b)-(ii) and (a,b)-(iii) respectively. The corresponding bifurcation diagram indicating the variation in $f_{3}(0)$ as a function of $L_{ee}/L$ is shown in (a)-(iv) and (b)-(iv). Additionally, for the specific case of the FF filament that is the focus of this paper, we find that beyond a critical value of compression (the value $L_{ee}/L$) that is a function of $A$, planar buckling shapes are not absolutely stable; the configurations stable to both in-plane and out-of-plane disturbances are three-dimensional twisted  shapes emerge. The onset of this secondary bifurcation governed by the torsional-to-bending stiffness ratio, $A$, shown in (a)-(iv). Note that the two dimensional shapes prior to the secondary bifurcation as well as twisted shapes post-secondary bifurcation are shown in (a)-(ii) for the specific instance $A=0.8$. The small oscillations in (a)-(iv) are artifacts arising due to the non-quasi-static nature of the simulation and the finite time used in the integration scheme.}
\label{fig:fig1}
\end{figure*}

\begin{figure}
\centering
\includegraphics[width=0.86\columnwidth]{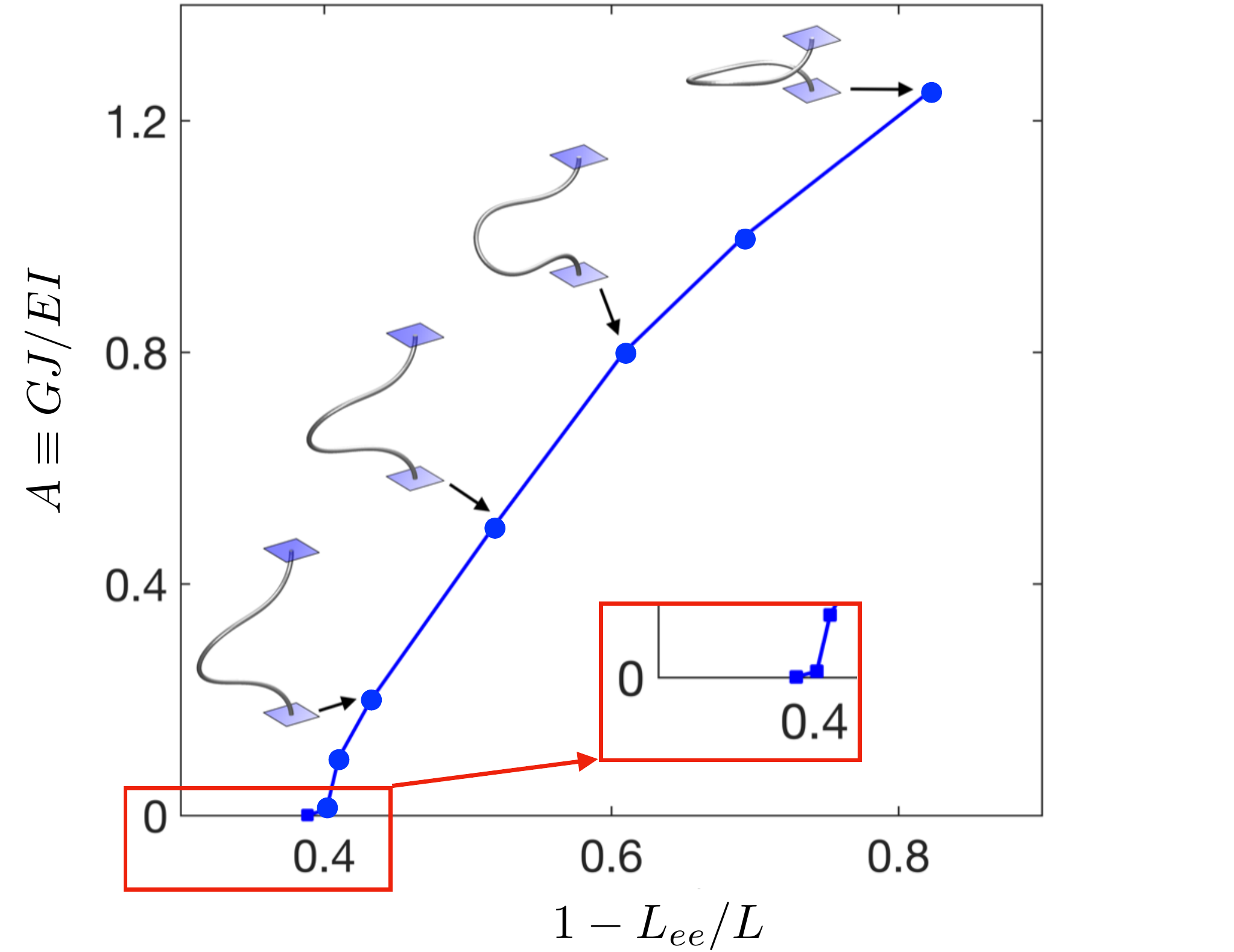}
\caption{For a slender rod with fixed-fixed (FF) boundary conditions, the ratio of torsional stiffness to bending stiffness, $A$, determines the slack $1-L_{ee}/L$ at the onset of secondary bifurcation. This in turn sets the value of the internal compressional stress field at bifurcation. The inset in red indicates the sensitivity of the bifurcation to $A$. }  
\label{fig:fig2}
\end{figure}

\begin{figure*}
\centering
\includegraphics[width=2\columnwidth]{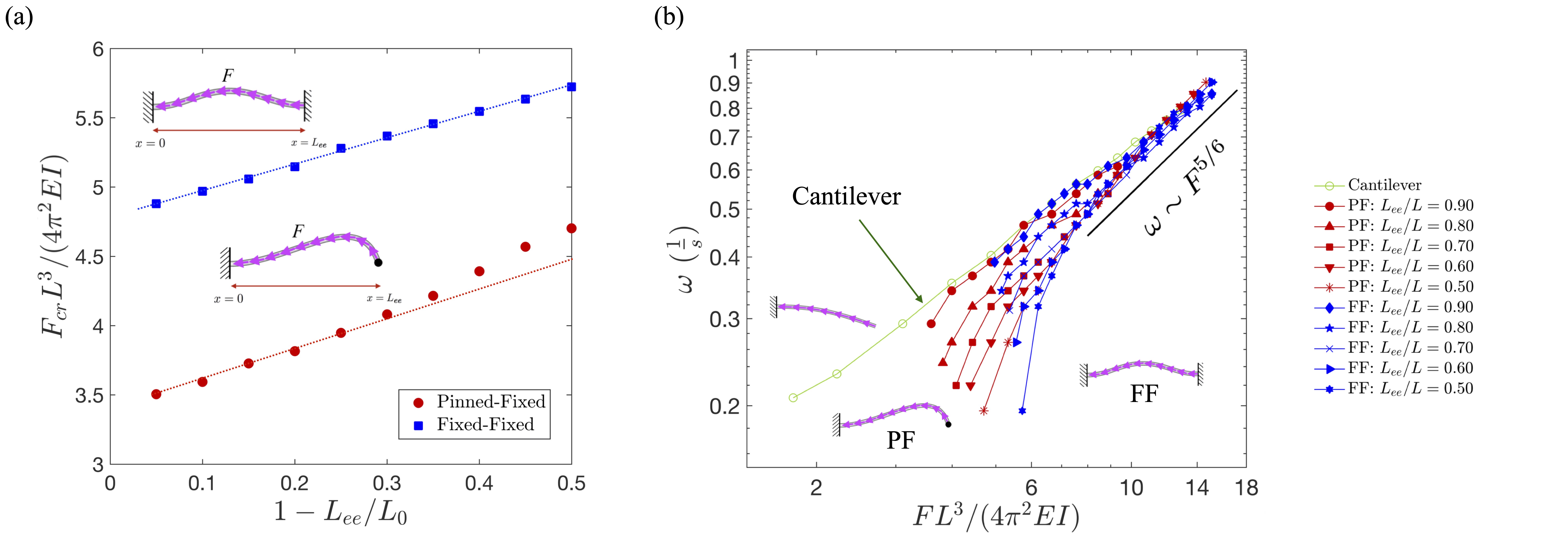}
\caption{Critical load for the onset of flapping, $F_{\mathrm{cr}}$ versus the compression rate, $1- L_{\mathrm{ee}}/L$ for both pinned-fixed (PF) and fixed-fixed (FF) scenarios vary with a linear relationship as shown in part (a). Here, the pink arrows schematically represent the follower forces and indicate the direction in which these are exerted. Frequency of the flapping oscillations are plotted in part (b) as a function of the force density, $F$ in logarithmic scales to illustrate two salient features; (1) as the follower force increases to values much larger than the critical limit, the effect of the pre-stress diminishes--far from criticality, similar frequencies are observed for all boundary conditions including the cantilever scenario--and (2) flapping frequencies in the limit $F \gg F_{\mathrm{cr}}$ scale roughly as $\omega \sim F^{5  \over 6}$ consistent with theoretical prediction \cite{soheil18a}.}
\label{fig:fig3}
\end{figure*}

\section{Computational scheme: Animated continuum Rod Model}

\subsection{Governing equations following the Kirchhoff approach}

The continuum rod model that we use follows the Kirchhoff's approach \cite{Kirchhoff} assuming each cross-section of the rod to be rigid and is described in detail elsewhere 
\cite{soheil18b}. Here we briefly summarize the model paying particular attention to the parameters that are varied in the investigation presented in this article. Note that we use the description slender rod and filament interchangeably in what follows. 

In this continuum-mechanics framework of the rod model, the  equilibrium equations (equations \ref{linear_momentum} and \ref{angular_momentum}) and the compatibility conditions (equations \ref{position_continuity} and \ref{orient_continuity}) are:
\begin{equation}
m(\frac{\partial \mathbf{v}}{\partial t} + \bm{\omega} \times \mathbf{v}) - (\frac{\partial \mathbf{f}}{\partial s} + \bm{\kappa} \times \mathbf{f}) - \mathbf{f_e} = \mathbf{0}, \label{linear_momentum} 
\end{equation}
\begin{equation}
\mathbf{I_m}\cdot \frac{\partial \bm{\omega}}{\partial t} + \bm{\omega} \times \mathbf{I_m} \cdot \bm{\omega} -(\frac{\partial \mathbf{q}}{\partial s} + \bm{\kappa} \times \mathbf{q}) + \mathbf{f} \times \mathbf{r} - \mathbf{q_e}= \mathbf{0}, \label{angular_momentum} 
\end{equation}
\begin{equation}
\frac{\partial \mathbf{r}}{\partial t} + \bm{\omega} \times \mathbf{r} - (\frac{\partial \mathbf{v}}{\partial s} + \bm{\kappa} \times \mathbf{v})  =\mathbf{0}, \label{position_continuity} 
\end{equation}
\begin{equation}
\frac{\partial \bm{\kappa}}{\partial t} - (\frac{\partial \bm{\omega}}{\partial s} + \bm{\kappa} \times \bm{\omega})  =\mathbf{0}.  \label{orient_continuity}
\end{equation}
Here $s$ parameterizes the arc-length of the rod and thus the location of material points along its backbone (centerline), $t$ is time, $m(s)$ is the mass per unit length (here set to a constant), and tensor $\mathbf{I_m}(s)$ is the moment of inertia per unit length. 

Equations (1)-(4) encapsulate both effects of geometry as well as the forcing driving the filament away from its straight base shape. Geometry dictates that the centerline tangent vector is $\mathbf{r}(s, t)$ and its variations along the length ($\partial \mathbf{r}/\partial s$) capture shear and extension. In this paper, such variations are assumed to be zero to ensure inextensibility and unshearability, therefore $\mathbf{r}$ becomes constant and collinear with the cross-sectional normal vector $\hat{\bf a}_3$ (see Figure 1 (a)-(i)). Vectors $\mathbf{f_e}$ and $\mathbf{q_e}$ are the external distributed force and moment, respectively. The spatial and the temporal derivatives in equations (\ref{linear_momentum}) - (\ref{orient_continuity}) are relative to the body-fixed frame 
$({\hat{\bf a}}_{1}, {\hat{\bf a}}_{2}, {\hat{\bf a}}_{3})$.

\subsection{External forces - active forces and passive drag}
In addition to internal forces and torques that act on each elemental section of the filament, we have externally exerted forces and torques on the system. These terms could be a consequence of boundary conditions imposed at the ends at $s=0$ and at $s=L$. Additionally, in the interior domain of the filament $s < 0 < L$, external forces acting on segments of the rod can arise from two independent physical processes or actions. 
The first one is the active animating force treated here as a distributed follower force that enters in (1) as the force density (per unit length) ${\bf f_{e}} $. This force  always acts along the local instantaneous tangent vector and is always directed (arc-length wise) towards the left fixed boundary $s=0$.  For a straight rod, these constitute a compressive force that initiates the buckling process. The active force thus pumps energy into the system.
The second type of external force we consider is dissipative fluid drag and serves to draw and extract energy away from the system. The specific form we consider is described later in this section. 

\subsection{Constitutive laws relating bending to curvature}

The unknown variables in equations (\ref{linear_momentum}) - (\ref{orient_continuity}) are: the vector $\bm{\kappa}(s,t)$ that captures two-axes bending and torsion, the vectors $\mathbf{v}(s,t)$ and $\bm{\omega}(s,t)$ that represent the translational and the angular velocities of each cross-section, respectively, and the vector $\mathbf{f}(s,t)$ that represent internal shear force and tension. For simplicity, we relate the internal moment vector $\mathbf{q}(s,t)$ in the angular momentum equation (\ref{angular_momentum}) to $\bm{\kappa}(s,t)$ through the linear constitutive law
\begin{equation}
\mathbf{q}(s,t) = {\mathbf{B}} \cdot \bm{\kappa},  \label{const}
\end{equation}
where ${\mathbf{B}}(s)$ represents the rod's bending and torsional stiffness. 

Without the loss of generality, we choose the body-fixed frame to coincide with the principal torsion-flexure axes of the rod, so that the stiffness tensor ${\mathbf{B}}$ for an isotropic rod can be expressed as a diagonal matrix 
\begin{eqnarray}
   [{\mathbf{B}}]=
  \left[ {\begin{array}{ccc}
    EI & 0 & 0 \\
    0 & EI & 0 \\
    0 & 0 & GJ\\
  \end{array} } \right].
  \label{stiffness}
\end{eqnarray}
In equation (\ref{stiffness}), $E$ is the Young's modulus, $G$ is the shear modulus, and $EI$ and $GJ$ are the bending and torsional stiffness respectively. The ratio of the two moduli 
\[
A \equiv {GJ \over EI}
\]
serves as an independent parameter that we choose to vary in our computations along with the magnitude of the active force, $F$.

In our simulations, we kept filament geometry and material properties such as density constant. Note that for a slender filament/rod with circular cross section and uniform properties, the moment of inertia $I$ and polar moment  of inertia $J$ are related by a constant factor independent of geometry. Furthermore $G$ and $E$ are related through the Poisson's ratio $\nu$ via $G = E/2(1 + \nu)$. In addition to computations using parameters listed in Table 1, additional calculations were conducted by varying the Poisson ratio. For constant filament geometry (radius and length), therefore this amounts to changing $A$ while keeping $I$ and $J$ fixed. 

\subsection{Fluid dissipation though Morison drag}
Finally, the dissipative environment and its interaction with the rod is modeled via the Morison drag form.
Note that in general the Morison drag equation is the sum of an inertia force in phase with the local flow acceleration (the functional form as found in potential flow theory) and a drag force proportional to the square of the relative far-field flow velocity (of the form for a rigid body placed in a steady flow). Here to calculate the drag force ${\bf f}_{\mathrm{M}}$, we choose to implement the equation as used in our earlier work \cite{goyal:05b, soheil18a}
\begin{equation}
{{- \mathbf{f}_{\textrm{M}}} \over {
\frac{1}{2}\rho_{\textrm{f}} d}}
= \Big( C_{\perp} |\mathbf{v}\times \mathbf{t}|\mathbf{t}\times(\mathbf{v}\times \mathbf{t}) + \pi C_{\|}(\mathbf{v}\cdot \mathbf{t})|\mathbf{v}\cdot \mathbf{t}|\:\mathbf{t} \Big).
\label{eq:dragM}
\end{equation}
Here, $\rho_{\textrm{f}}$ is the fluid density, $d$ is diameter of the rod, ${\bf t}$ is the unit tangent vector to centerline and $C_{\perp}$ and $C_{\|}$ are the normal and tangential components of drag coefficients, respectively. 

The Morison formula captures the hydrodynamic resistive forces applied to cylindrical structures in oscillatory or steady flows \cite{FISH198015,WOLFRAM1999311} reasonably well. More generally, the local drag force calculated by Morison equation as in (7) is scaled by drag coefficients that depend on Reynolds number and surface roughness. 

Note that the slender rod considered here has weak inertia. In previous work \cite{soheil18a}, we investigated the effect of both linear and quadratic (Morison) drag on planar flapping of clamped rods. While the form of the drag force has a quantitative effect on the amplitude and frequency of the flapping oscillations, we found that once oscillations were initiated, the qualitative response  was similar. Here we have chosen to focus on the quadratic form of the drag that provides enhanced damping and allows for extrapolation of our results to macro-scale structures where the low Reynolds number criterion crucial to the establishment of Stokes drag may not be valid. 

Additionally,  
our previous work \cite{soheil18a} - in the context of purely two dimensional buckling instabilities with disturbances also restricted to planar forms - suggests that the critical point at which oscillations occur are nearly independent of form of drag, and  drag and inertia together may determine the onset of instability. However, for large follower force densities ($F$ large), fluid drag dominates the dynamical properties of the stable states. In this work, keeping the densities constant,  we focus on the influence of elasticity and activity on the selection and characteristics of fully non-linear, stable deformations attained by the rod. Analyzing the critical point and the identification of the critical value of the active force below which oscillations may not occur is left to future work. 

\begin{table}[!h]
\begin{center}
   \begin{tabular}{lcc}
    { \bf Quantity} & {\bf Symbol} & {\bf Value}  \\\\ \hline
    \hline
    \\
    Bending Stiffness & $EI$ &  29.231 N m$^2$\\ 
    Torsional Stiffness & $GJ$ &  23.385$^{*}$ N m$^2$\\ 
    Mass per unit length& $m$ & 0.2019  kg/m \\ 
    Length & $L$ & 8  m \\ 
    Diameter & $d$ & 0.0096  m \\ 
   \\
    \hline
    \\
    Normal drag coefficient & $C_{\perp}$ & 0.1  \\ 
    Tangential drag coefficient & $C_{\|}$ & 0.01   \\ 
    Surrounding fluid density & $\rho_{\textrm{f}} $ & 1000  kg/m$^3$  \\ 
    \\
    \hline \hline
    \end{tabular}
    \caption{Representative numerical values for the properties of the rod, and for the drag coefficients used in the simulations. The density of the surrounding fluid is taken to be the density of water. The slender rod is comprised of material that is {\em not} massless. The densities are kept constant for all computations. $^{*}$Note that $G$ and $E$ are related through the Poisson ratio $\nu$ via $G = E/2(1 + \nu)$. Additional computations with varying ratios $A \equiv EI/GJ$ were further conducted  using different values of $G/E$ obtained by varying Poisson ratio. For a slender filament/rod with circular cross section and uniform properties, the ratio $I/J$ of the moment of inertia $I$ and polar moment  of inertia $J$, are related by a constant factor independent of geometry.}
    \label{tab1}
\end{center}
\end{table}

\section{Results: Computational Simulations  and Discussion}

\subsection{Solution methodology, accuracy and convergence}

The \textit{Generalized-$\alpha$} method is adopted to compute
the numerical solution of this system, subjected to necessary
and sufficient initial and boundary conditions. A detailed description of this numerical scheme applied to the rod formulation is given in \cite{soheil18b} and references therein. Our computational model has been validated by comparing the critical value of the follower force in the planar cantilever scenario with analytical results of Beck's column \cite{soheil18a}. In this paper, we analyze the spatiotemporal response and long-time stable dynamical states attained by pre-stressed rods with fixed-fixed (FF) and pinned-fixed (PF) boundary conditions subject to and animated by a uniformly distributed follower force. 

In all the computations, an initially straight cylindrical rod is used with the properties given in Table \ref{tab1} chosen to represent a soft filament. Before applying the distributed follower force, the pre-stress is generated by axial compression of the rod that leads to buckled equilibrium shapes as shown in Figure \ref{fig:fig1}. The pre-stress is specified in terms of the end-to-end distance, $L_{ee}$. The buckled shapes can be planar or out-of-plane and are described next, before exploring the effects of follower force.

\subsection{Two and three dimensional buckled shapes in the absence of follower force}

Figure \ref{fig:fig1} (a) and (b) show the post-buckling equilibria for fixed-fixed (FF) and pinned-fixed (PF) boundary conditions, respectively. For both the boundary conditions, the buckling onsets in plane as the compressive load $f_3(0)$ reaches Euler buckling load, which is  $P_{\mathrm{cr}}=4\pi EI/L^2$ for fixed-fixed (FF) and $P_{\mathrm{cr}}=2.045\pi EI/L^2$ for pinned-fixed (PF) conditions. The compressive load $f_3(0)$ initially increases with the pre-stress $1-L_{ee}/L$ in each case. Supplementary video ESM Movie 1 shows the quasi-static simulation of how the rod shapes evolve for $A=0.8$.

For the fixed-fixed (FF) case, however, which is torsionally constrained, once compression increases beyond a critical limit (for example, $1-L_{ee}/L \approx 0.60$ for $A=0.8$) planar buckled shapes become unstable and rod transitions to energetically more favorable out-of-plane configurations that allow for the partial storage of strain energy in torsion \cite{goyal:05b,VANDERHEIJDEN2003161}. This critical limit of secondary bifurcation increases as the torsional-to-bending stiffness ratio, $A=GJ/EI$ increases, as depicted in the bifurcation diagram of Figure \ref{fig:fig1} (a)-(iv). Figure 1(a)-(ii) for instance shows both the stable two dimensional shapes (for pre-stress less than that required for secondary bifurcation) as well as the fully twisted, three-dimensional stable shapes attained post-secondary bifurcation.  

In the pinned-fixed (PF) case, by contrast, which is not torsionally constrained, no secondary bifurcation emerges by increasing the end-to-end compression. Supplementary material ESM Movie 2 demonstrates quasi-static simulation used to generate the results for the equilibria shown in Figure \ref{fig:fig1} (b). Note that figure 1(b)-(iv) shows the bifurcation diagram for the PF case with figure 1(b)-(ii) pictorially depicting the two dimensional shapes that are determined via computation.

In the following sections we take a representative variety of stable buckled configurations as \textit{base states} for exploring the dynamics under distributed follower forces with homogeneous intensity. In particular, we first focus on the range of end-to-end values with planar base states in both FF and PF scenarios for $A=0.8$, i.e., the interval between points 1 and 2 in Figure \ref{fig:fig1} bottom row ($0.1\leqslant 1 - L_{ee}/L\leqslant 0.5$). Next, we focus on the out-of-plane buckled configurations of FF scenario as base states. We also investigate the effect of A as well as overall material stiffness. The role of $A$ is crucial as seen in the difference in the extent to which the three-dimensional base states curve and bend out of plane - illustrated in Figure 2.

\subsection{Flapping Motion of Planar Base States}

In this section we examine how the level of pre-stress (through the imposed value of $L_{ee}/L$)  and the magnitude of the active force density $F$ control (1) the stability of base states,  and (2) the frequency of emergent oscillations. In order to generate the appropriate boundary conditions, we envision the rod/filament 
to be constrained at the ends via two clamps (for the FF case), or with a pin joint and a clamp (for the PF case). By controlling the end-to-end distance $L_{ee}$, we can then first generate stable buckled shapes, and then subsequently apply a follower force of intensity $F$ along the local tangential direction of the rod's centerline. 

Keeping the torsional-to-bending stiffness ratio constant at $A=0.8$, for near the entire range of end-to-end values within the planar buckling regime, i.e. $0.05\leqslant 1 - L_{ee}/L\leqslant 0.5$ our numerical analysis shows that beyond a critical value of the follower force, $F_{\mathrm{cr}}$, buckled shapes no longer maintain static equilibrium and planar oscillations - flapping oscillations - emerge when rod is subject to any infinitesimal {\em planar}  perturbations. This result is consistent with the onset of Hopf bifurcation that is obtained by linear stability analysis in cantilever scenario \cite{PMID:24352670,DeCanio20170491,soheil18a,Arvind20a} 
ESM Movie 3 in the included supplementary material demonstrates an example of flapping oscillations with the spatiotemporal distribution of curvature and angular velocity in the fixed-fixed scenario.

Our analysis reveals that in both fixed-fixed and pinned-fixed scenarios:
1) $F_{cr}$ increases linearly with pre-stress - see Figure \ref{fig:fig3}(a),
2) near the critical point, the frequency of steady oscillations are sensitive to end-to-end distance (Figure \ref{fig:fig3}(b)), 
3) and far from the critical point (for large values of $F$) the frequency of oscillations become independent of the pre-stress and scale roughly as $\omega \sim F^{5 \over 6}$. The origin of the power law exponent ${5 \over 6}$ is the balance between the active energy per unit cycle pumped into the system by the follower force that is dissipated away by the fluid via the quadratic Morison drag form. A different form for the fluid drag will lead to a different coefficient as we have pointed out earlier \cite{soheil18a}.

We observe from  Figure \ref{fig:fig3}(b), that for very large values of $F$, the nature of the boundary conditions seem to play a diminishing role in setting the value of the flapping frequency. This may be rationalized as follows. Far from the critical point, such as when scaled follower force becomes one order of magnitude larger than the critical bifurcation force ($FL^3/(4\pi^2EI)\approx10$), the effect of the pre-stress becomes dominated by (negligible with respect to) the relatively large forces induced by the active force, $F$. 

Interestingly, we find that for the pinned-fixed scenario, the existence of oscillations depends on the direction of the follower forces. Specifically, when the follower force is directed from the clamped end towards the pinned end, no dynamical instability is induced \cite{2019arXiv190508421F}.  This results suggests that altering boundary conditions independently of the strength of the follower forces or mechanisms generating these can be used to initiate or quench flapping oscillations.

Finally, we would like to point out that we do not investigate the onset of oscillations. In the system described by equations (1)-(6), fluid drag, active forces, rod elasticity and the inertia of the rod all enter into the picture. In earlier work on cantilever rods \cite{soheil18a}, we used a simple model to studied how each of these influences the onset of oscillations and the location of the critical point. For the pre-stressed rods with inertia considered here, the critical points are not easily identifiable using time stepper techniques such as our model.  Hence we focus on the fully non-linear and large amplitude solutions far from onset.

\begin{figure*}
\centering
\includegraphics[width=2\columnwidth]{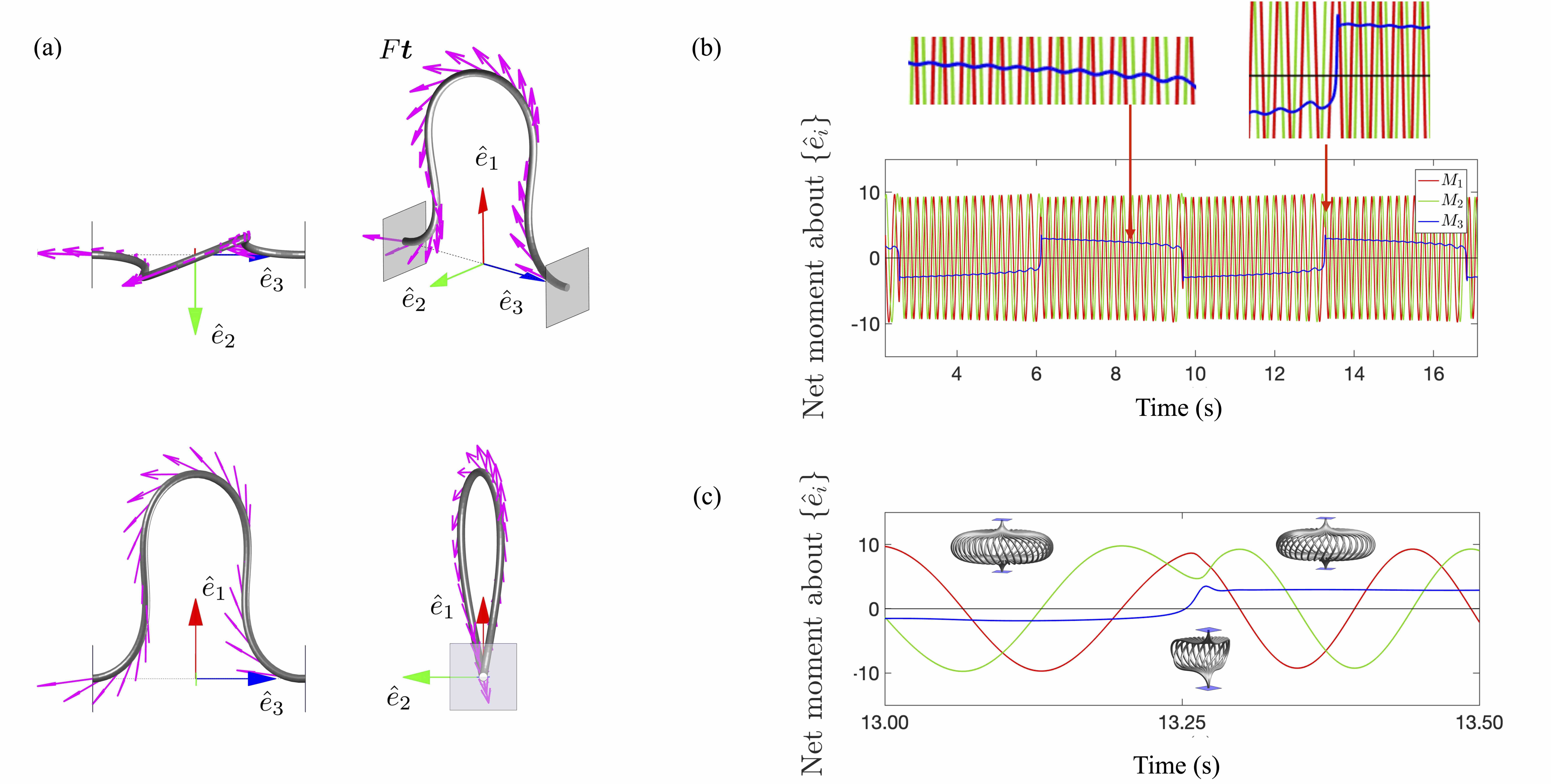}
\caption{Three orthogonal views and a perspective view of the rod corresponding to a base state (stable, static buckled equilibria) with $1-L_{ee}/L=0.6$ and $A=0.8$ are shown in part (a) in {\em{3rd angle projection}} of engineering convention. The red, green and blue arrows represent the global reference frame in each view. 
When this twisted rod is now subject to  distributed active follower force, $F\mathbf{t}$, acting along the tangent vector and shown here as magenta colored arrows, swirling oscillations about the end-to-end axis, $\hat{e}_3$ emerge. The direction of swirling motion periodically flips along this axis. Net moment of the follower force $\mathbf{M}=(M_1,M_2,M_3)$ is calculated by the formula $\int^L_0 (\mathbf{R_c} \times F\mathbf{t}) ds$ where $\mathbf{R_c}$ is the position vector of the cross-section at $s$ with respect to the inertial frame $\{ \hat{\bf e}_i \}$ located at the midpoint between the two clamps. The component of the moment generated about the end-to-end axis, $M_3$, is shown with blue curve in parts (b) and (c), and is the main driver of the rotational swirling. On the other hand, moments generated about $\hat{\bf e}_1$ and $\hat{\bf e}_2$, i.e. $M_2$ and $M_3$ shown respectively with red and green, have a phase difference of $\pi/2$, and their effect is to reduce the torsional deflection towards zero and increase the bending deflection. This process of torsional reduction and bending enhancement driven by follower forces continue until the shape of the rod becomes nearly planar; subsequent to this a flip occurs
and the swirling direction reverses.}  
\label{fig:fig4}
\end{figure*}

\begin{figure*}
\centering
\includegraphics[width=2\columnwidth]{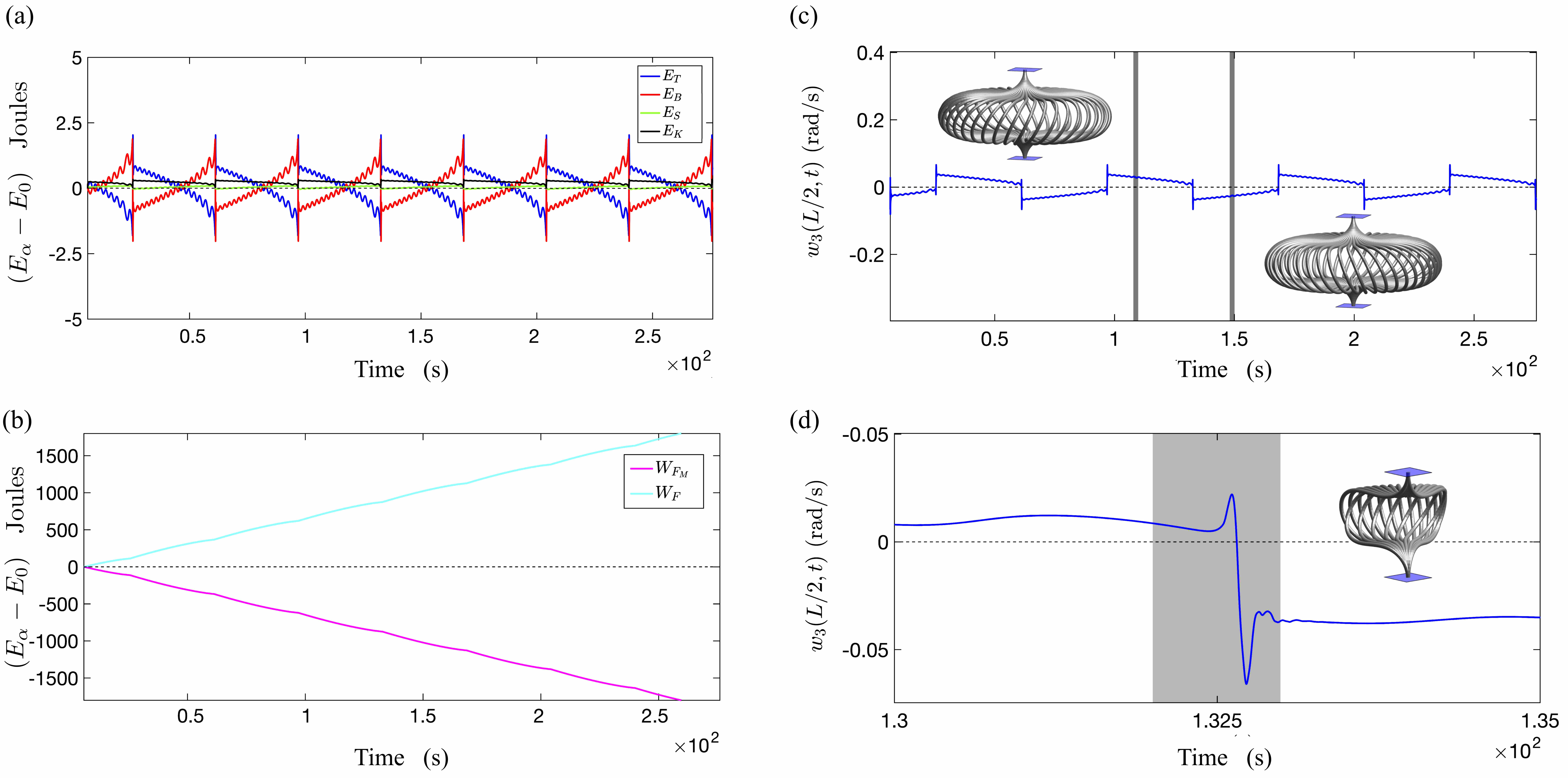}
\caption{Variations in time of torsional energy, $E_T$, bending energy, $E_B$, total strain energy, $E_S$, and kinetic energy, $E_K$ with respect to the energy levels at the base state, $E_0$ are shown in part (a) and variations of energy dissipated by fluid drag, $W_{\mathbf{f}_{\textrm{M}}}$, and work done by follower force, $W_F$ are depicted in part (b). The results reveal that a relatively small change occurs in both the strain energy (the green curve) and the kinetic energy (the black curve), while the torsional energy (blue curve) falls bellow its corresponding magnitude at the base state ($E_0$) and the bending energy increases to significantly larger values compared to the based state. As a measure of the swirling frequency, the component of the angular velocity about $\hat{a}_3$ at the mid-span length is plotted in time for $|F|=1$ N/m, $1- L_{ee}/L=0.60$ and $A=0.8$ in part (c). A set of superimposed shapes during an entire swirling cycle (the shaded interval) are also given in parts (c) and (d). Evolution of angular velocity at the mid-span length during the flipping along with the rendition of the shapes is illustrated in part (d).}
\label{fig:energy}
\end{figure*}

\begin{figure}
\centering
\includegraphics[width=1\columnwidth]{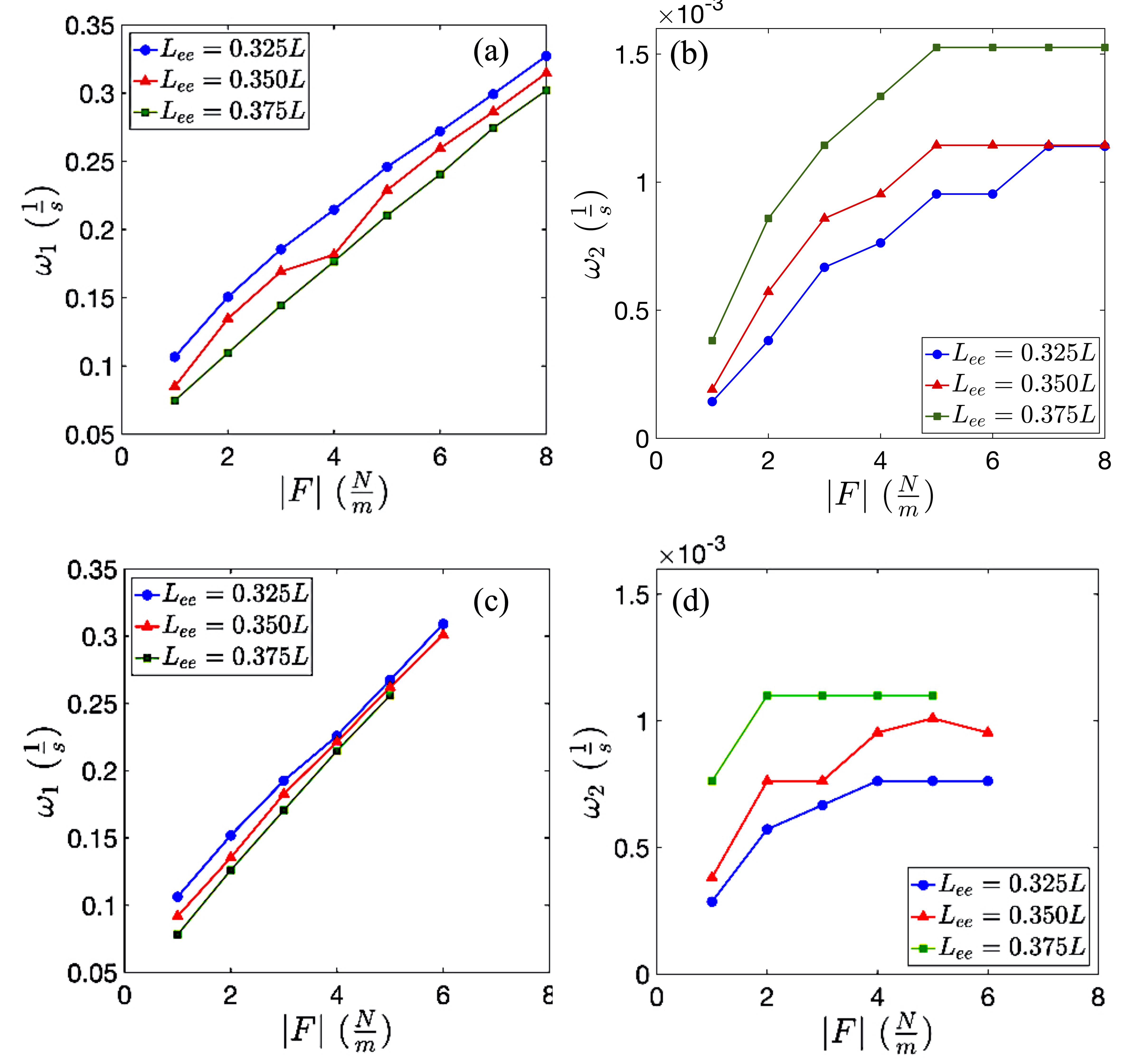}
\caption{The swirling frequency and the rate of flipping as a function of follower force density in the interval $0.325\leqslant L_{ee}/L\leqslant0.375$ and for $A=0.8$ are shown in parts (a) and (b), respectively. The results reveal that for a fixed value of follower force by enhancing the end-to-end compression (reducing $L_{ee}$) the swirling frequency, $\omega_1$ increases, while the rate of flipping, $\omega_2$ decreases. We also find that $\omega_1$ increases roughly linearly with the follower force intensity (over the range investigated), but $\omega_2$ initially increases and then becomes insensitive to the force intensity. The results for a rod with the same $A$ value but with bending and torsional stiffness $EI$ and $GJ$ each twice as small as before, is also shown in parts (c) and (d). For this softer filament we find frequency of rotations to be more sensitive to the follower force and less sensitive to the compression $L_{ee}$.
}
\label{fig:fig6}
\end{figure}

\subsection{Swirling Motion with Periodic Reversal of Non-Planar Base States}

For the range of parameters with non-planar base shapes tested here ($0.60\leqslant1-L_{ee}/L\leqslant0.675$ and $A=0.8$ and $A=1.0$) we find purely rotational oscillations to arise when a non-zero follower force is applied to the rod. We call this \textit{swirling} motion. Figure \ref{fig:fig4} shows an example of the base states in this compression range. By keeping the force constant and letting the dynamics to reach a steady state, we find that swirling rates decrease gradually until an abrupt reversal of direction - or \textit{flipping} - occurs (see Figures \ref{fig:fig4} (c,d) and \ref{fig:energy} (c,d) and ESM Movie 4). Therefore, from simulations we extract a second characteristic time-scale - the rate at which flips are observed. 

Figure \ref{fig:fig4} (c,d) show computed moments generated by the follower forces measured about a coordinate system located at the midpoint of the end-to-end axis. We hypothesize that the component of this moment about the end-to-end axis, $\hat{\bf e}_3$ serves to drive the rotational motion - i.e, swirling. On the other hand, the effect of moments generated about $\hat{\bf e}_1$ and $\hat{\bf e}_2$ (which seem to be out of phase by $\pi/2$) is to diminish the torsional deflection towards zero and increase the bending deflection. This reduction in torsional deflection is accompanied by the attainment of near planar configurations by the rod. At this point,  the rod is observed to then snap again into a highly twisted shape in the opposite direction. 
 
In each flipping cycle which - for this particular value of $L_{ee}$ and $A$ - takes about two orders of magnitude longer than a swirling cycle, the total strain energy slightly increases (less than $0.1\%$ per cycle for the results shown in Figure \ref{fig:energy}); the bending energy however increases significantly (about $4\%$ per cycle) while the torsional energy decreases significantly (about $100\%$ per cycle).
To better describe this phenomenon from a mechanical point of view, in the following segment we examine energy variation relative to the energy of the base states.

\subsection{Energy Exchange During Oscillations}

Expressions used to evaluate the torsional energy, $E_T$, bending energy, $E_B$, total strain energy, $E_S$, kinetic energy, $E_K$, energy dissipated by fluid drag, $W_{\mathbf{f}_{\textrm{M}}}$, and work done by follower force, $W_F$ from simulation results are given below, respectively.
\begin{equation}
\begin{split}
E_s &=\frac{1}{2} \int_0^L (\mathbf{B} \cdot \bm{\kappa} )\cdot \bm{\kappa}\: ds = E_B+E_T \\
&=\frac{1}{2} \int_0^L EI (\kappa_1^2 + \kappa_2^2) \: ds + \frac{1}{2} \int_0^L GJ \kappa_3^2 \: ds 
\label{eq:straine}
\end{split}
\end{equation}
\begin{equation}
E_K=\frac{1}{2} \int_0^L m(\mathbf{v} \cdot \mathbf{v} )\:ds
\label{eq:kinetice}
\end{equation}
\begin{equation}
W_{\mathbf{f}_{\textrm{M}}}=\int_0^t \int_0^L (\mathbf{f}_{\textrm{M}} \cdot \mathbf{v}) \: ds \: d\tau
\label{eq:fluide}
\end{equation}
\begin{equation}
W_F=\int_0^t \int_0^L (F\mathbf{t} \cdot \mathbf{v}) \: ds \:d\tau
\label{eq:followere}
\end{equation}

Figure \ref{fig:energy} (a) demonstrates various forms of energy relative to the energy levels of the base state configuration after a unit follower force is applied to a buckled rod with $1- L_{ee}/L=0.60$ and $A=0.8$. Once the force is applied and rotational (swirling) cycles begin, we observe a gradual decrease in kinetic energy and a gradual increase in total strain energy--see Figure \ref{fig:energy} (a,c). With further swirling, when enough of torsional energy is converted to bending, rod reaches a nearly planar configuration that is not stable. This triggers a change in direction of the swirling rotations through which high levels of bending energy are discharged back into torsional form--see Figure \ref{fig:energy} (d). The discharge of bending energy back into the torsional form also contributes to a small jump in the kinetic energy that passes through zero during the moment of reversal. Flipping thus involves the relaxation (or drop) of both bending and total strain energy. 

Figure \ref{fig:energy} (b) illustrates that work done by the follower force, $W_F$ has almost the same rate of change and magnitude as the energy dissipated in the fluid medium, $W_{\mathbf{f}_{\textrm{M}}}$. This suggests that a much smaller proportion of the work done by active forces is stored in the system as strain energy. However, the constrained configuration allows for a periodic transfer of entire elastic energy in the torsional mode into and from bending, thus resulting in a dynamical trajectory with two distinct time scales of swirling and flipping. 

\subsection{Frequencies of Oscillations - Swirling and flipping frequencies}

This section examines the sensitivity of steady state oscillations to both pre-stress and force intensity. Figure \ref{fig:fig6} top row, shows the rates of both swirling and flipping dynamics as a function of follower force intensity and pre-stress. The results belong to three distinct pre-stress values, $L_{ee}/L=\{0.325,0.350,0.375\}$. In all cases, we find that swirling frequency, $\omega_1$ linearly varies with follower force intensity, while having a low sensitivity to the pre-stress. On the other hand, the flipping frequency, $\omega_2$ shown in the same figure is found to be more sensitive to the pre-stress values and becomes approximately independent of the follower force for very large active force densities. 

We find that the rate of flipping seems to saturate to a value that depends on the initial compression ratio $L_{ee}/L$. More generally, our computations suggest that 
swirling rates seem very sensitive to the follower force and drag while flipping rates are dominantly determined by the interplay of bending and torsion. Keeping $F$ constant while varying the initial level of pre-stress results in an increase in the swirling rate but a decrease in the flipping rate.

\subsection{Role of elasticity parameters $EI$ and $GJ$}

\begin{figure}
\centering
\includegraphics[width=0.8\columnwidth]{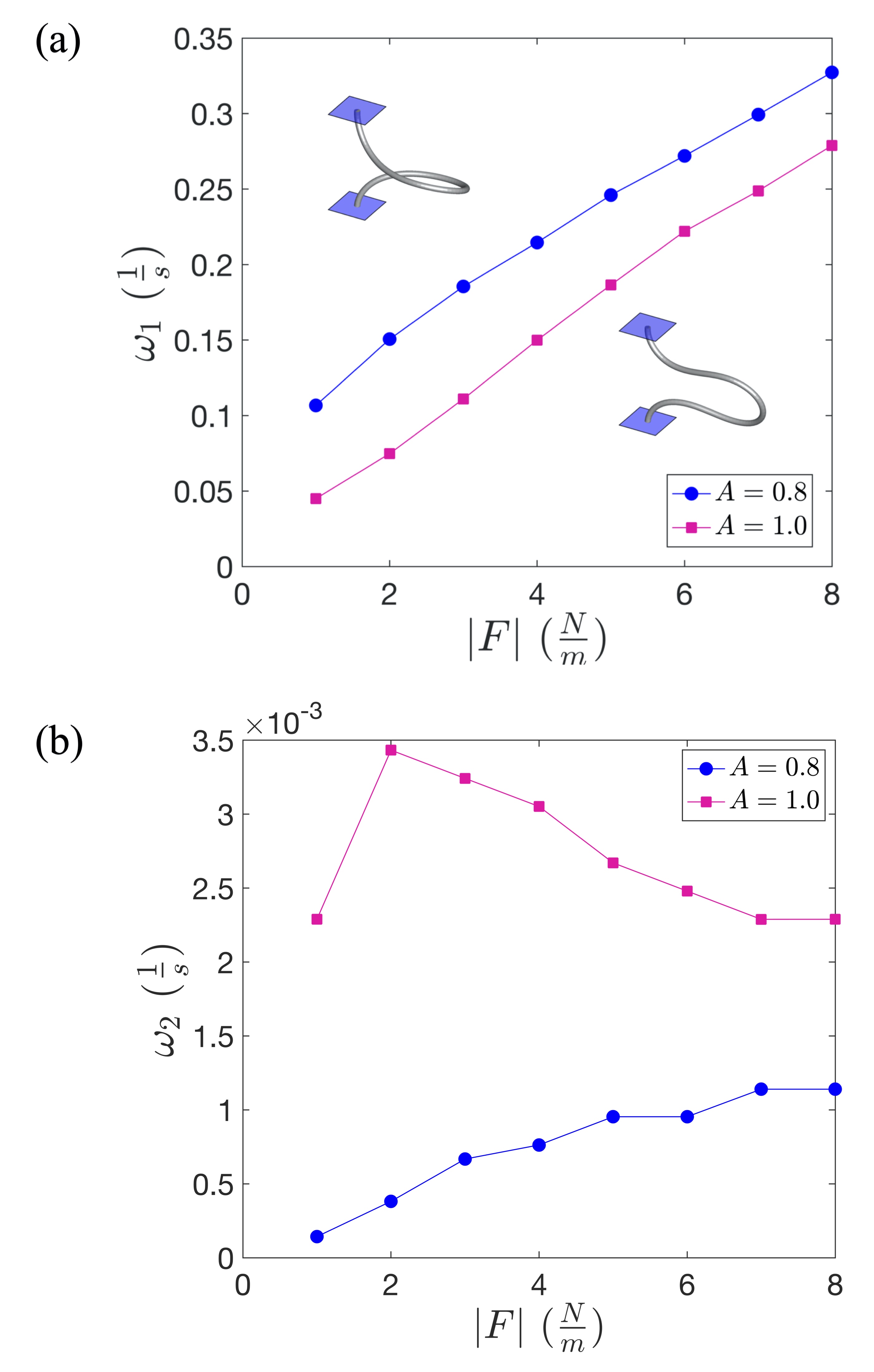}
\caption{Frequency of oscillations as a function of follower force density is compared for torsional-to-bending stiffness ratios varying from $A=0.8$ to $A=1.0$ (by keeping the bending stiffness constant at the value given in Table \ref{tab1}, and increasing the torsional stiffness, $GJ$), and for pre-stress level of $L_{ee}/L=0.325$. Part (a) shows that by increasing $A$ (with $F$ fixed), the torsional deflection of the base states decreases and so do the swirling frequencies $\omega_{1}$. In contrast, we find that for fixed $F$, the frequency of flips is higher for the larger value of $A$ (1.0 compared to 0.8). Note also, the intriguing result that for $A=0.8$, the flipping rate increases with $F$. At the higher value of $A=1$ however, we observe a non-monotonic regime with an increase following by a steady decrease in $\omega_{2}$ with $F$.}
\label{fig:fig7}
\end{figure}

\begin{figure*}
\centering
\includegraphics[width=1.86\columnwidth]{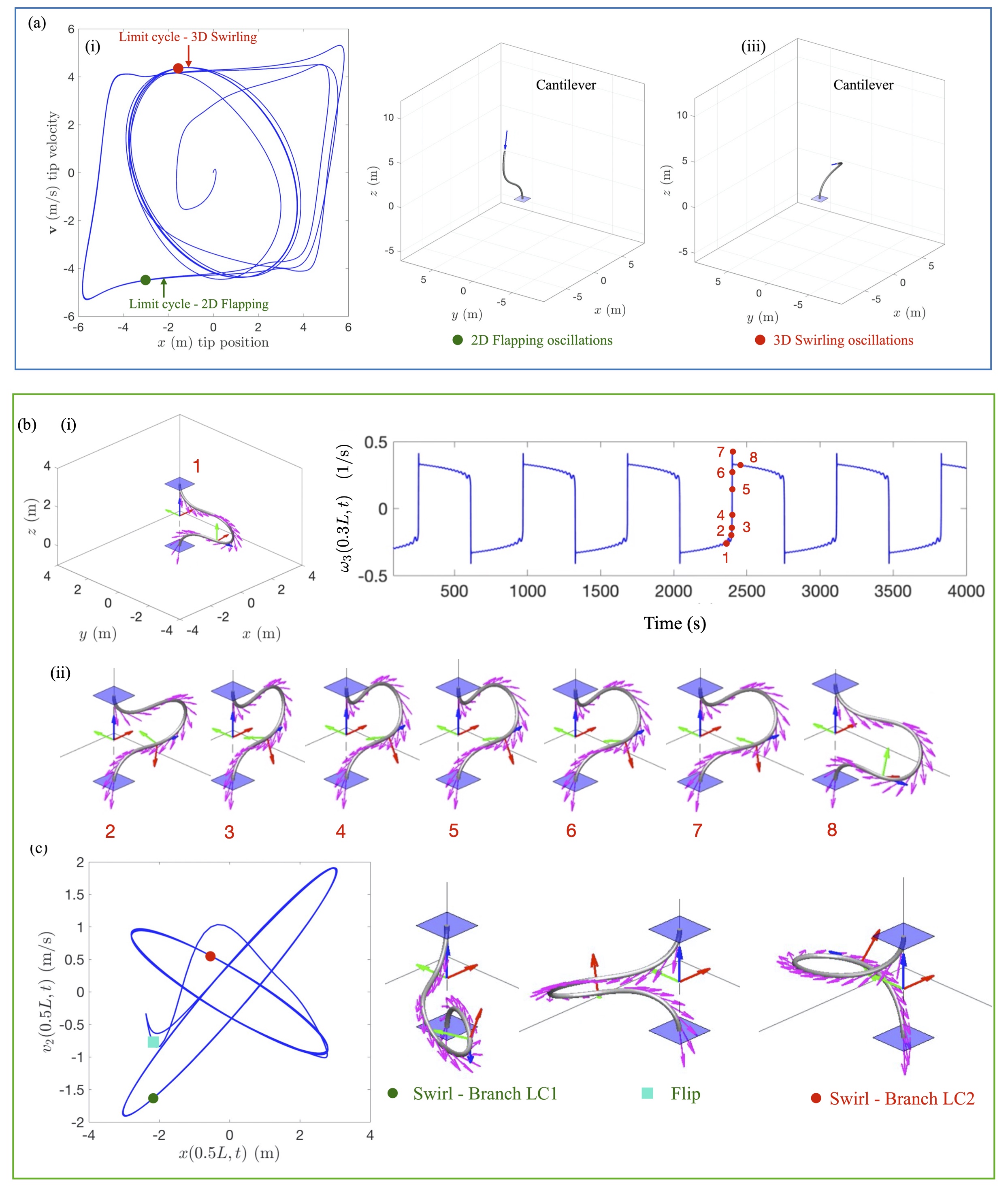}
\caption{Limit cycles for oscillatory instabilities: (a) active cantilever rods,  and (b,c) fixed-fixed pre-stressed rods. [Tile (a)] Results for a partially constrained cantilevered rod with no pre-stress subject to follower force (see also ESM Movie 6) when $f_3(L)=-20.2\pi^2 EI/4L^2$. (i, ii) We show the limit cycle and a typical shape for flapping modes (green). The limit cycle in (i) features the velocity of the free end of the tip as a function of its vertical height. (ii) Here, only planar disturbances are imposed. (iii) When three dimensional disturbances are allowed rod oscillations are fully three dimensional with continuous swirling - see limit cycle in (i), and snapshot in (iii).
(b) Here we show the sequence of shapes (time instants - 1 to 8,
see ESM Movie 5) for the swirl-flip transitions. The parameters are $F=4$, $L_{ee}/L = 0.4$, and $A = 0.8$.
Note the overshoot seen between numbers 6 and 8. (c) Illustration of the swirl-flip limit cycles for the same parameters as in (b). 
We show only a few cycles immediately before (when the dynamics corresponds to branch LC1 and after the swirl (when the dynamics correspond to branch LC2). The limit cycles are not absolutely coincident due to the gradual changes in energy described in section III E.
}
\label{fig:fig8}
\end{figure*}

To understand the effect of elasticity, which is captured in our model with parameters $EI$ and $GJ$ representing bending and torsional stiffness respectively, we keep all other properties constant and examine the dynamics of a filament with bending and torsional stiffness twice as smaller as the values reported in Table \ref{tab1}. Figure \ref{fig:fig6} (c,d), show the results for pre-stress rates belonging to $L_{ee}/L=\{0.325,0.350,0.375\}$. For the softer filament with smaller elastic properties we find the relationship between the rotational frequency and the follower force to be linear although with a larger slope compared to the stiffer rod shown in Figure \ref{fig:fig6} (a,b). This can be explained by the fact that rotational oscillations in this regime emerge by the interplay of (1) active energy entering the system, (2) elastic energy of bending and torsion, and (3) energy dissipated due to fluid viscosity. During rotational cycles in between flapping events, part of the active energy must be spent on increasing the strain energy of the system (see green curve, $E_s$ in Figure \ref{fig:energy} part (a.i)). For the softer filament this strain energy barrier is smaller, hence a larger portion of active energy remains available to overcome fluid resistance and to gain rotational momentum.

For the flipping frequency we find a response pattern similar to the filament with larger stiffness, however the saturation frequencies, i.e., the rate at which flipping rates become independent of the follower force, are smaller for softer filaments. Moreover, for the softer filament we observe a diminished sensitivity of both swirling and flipping rates to the pre-stress in comparison to the stiffer rod. This is due to the fact that for the softer filament restoring effects at the base states (e.g., $P_{cr}^{FF}=4\pi^2 EI/L^2$) are relatively smaller too.

We then examine the role of torsional-to-bending stiffness ratio, $A$, on swirling and flapping. For that, we keep the bending stiffness, $EI$, constant (at the value given in Table \ref{tab1}) and increase the torsional stiffness, $GJ$ to enlarge $A$ from $0.8$ to $1.0$. As shown in Figure \ref{fig:fig7}, the base state with $A=1.0$ has much smaller torsional deflection compared to the rod with $A=0.8$ for the same compression level of $L_{ee}/L=0.325$. When a follower force is applied to these two base states we find that larger torsional-to-bending stiffness ratio results in smaller swirling frequencies, which nonetheless vary with a similar slope against force density. 

The rod with larger $A$ exhibits smaller out-of-plane deflection, when all else is kept constant, and thus the follower forces generate smaller net moments that in turn this gives raise to smaller swirling frequencies, compared to the application of same forces to the rod with smaller $A$ and larger out-of-plane deflection. 

While the difference in the functional dependence of $\omega_{2}$ on $F$ for $A=0.8$ and $A=1.0$ evident from Figure \ref{fig:fig7}(b) may be surprising, a possible explanation can be recognized by considering more closely the value of $L_{ee}/L = 0.325$ and the distance the base state is located from the secondary bifurcation point. The inset is a close-up of Figure  \ref{fig:fig1}(a)-(iv). Since $1-L_{ee}/L = 0.625$ which for $A=1.0$ is closer to the secondary bifurcation point than $A=0.8$, this suggests that the absolute values of $L_{ee}/L$ and the value of $A$ control the frequencies; however implicit in this control is the shape of the base state and the energies stored in this state that depends on the distance from the secondary bifurcation.

Synthesizing our observations together, we find that the swirl-flip combination - the periodic changes in direction of the swirling rotations - is intricately associated with a specific sequence of energy interchanges between bending and torsional deformation modes of the filament. The non-conservative and configuration  slaved dependence of the follower force direction succeed in mediating this transfer even as the magnitude of the follower force is fixed. In this respect, our system may also be considered as generating relaxation oscillations due to the effect of periodic excitation forces; except that here the frequency of this excitation is not imposed but emergent. 

We deduce that the swirl-flip-swirl sequence is a self-generated emergent oscillation controlled intrinsically by the fact that an initially planar filament is unbiased insofar as the direction of swirling is concerned {\and} by the fact that the clamped-clamped filament is torsionally constrained. Indeed, relaxing the constraint at one end by converting the boundary conditions to a clamped-free cantilever type and approximating distributed follower forces by a single follower force exerted at the free tip (see Figure 8 and ESM Movie 6) results in the loss of the flips and continuous swirling. 

\section{Perspectives: Connections to relaxation oscillations}

The results presented in Section III for the clamped-clamped filament combined with results for the continuous swirling of an cantilevered active rod (Figure 8(a), with no torsional constraints at $s=0$) suggests that flipping  involves the relaxation (or periodic and slow variations) of both bending and total strain energy. The discharge of bending energy back into the torsional form also contributes to a small jump in the kinetic energy that passes through zero during the moment of reversal. The role of filament inertia - as included in the time derivative terms in equations (1) and (2) - in perhaps enabling overshoot and thus a change in the sign of the rotation direction  is unclear and requires future exploration. 

In the absence of active forces, fluid drag competes with physical mechanisms that temporarily store energy and produced damped motions. One of us has previously worked on theoretical and experimental systems with solid-fluid interactions where bulk solid elasticity provides the storage mechanism \cite{Arvind20a},  as well as in purely fluid-fluid contexts with surface tension serving to store energy temporarily \cite{Gopinath2001, Gopinath2002}. Here in the active context, fluid drag plays a crucial dual role - both dissipating the energy and  providing a pathway to stabilize the system by forcing the emergence of oscillations with large amplitude and clear frequencies. The dependence of the frequencies on the active force density $F$ follows power laws as shown in previous theoretical work  by us \cite{soheil18a,Arvind20a}); the exact exponent depends (provided one is far from onset) on the form of the drag and is $5/6$ for quadratic drag as shown here, and $4/3$
for linear drag forms such as low Reynolds number Stokes drag.

We can make further connections to the classical relaxation oscillations  eponymous with Balthazar van der Pol's paper \cite{doi:10.1002/9783527617586.ch1} studied in relation to self-sustaining nonlinear oscillations in triode circuits. 
Self-oscillating relaxation oscillations have been observed in electromagnetic devices \cite{doi:10.1002/047134608X.W2282}, semiconductor laser devices subject to feedback concomitantly with optical injection \cite{10.1117/12.662832}, in the context of acoustics and music \cite{JENKINS2013167}, and more recently found ubiquitously in cell and system biology contexts (see \cite{cite-keyElowitz,cite-key_cc} and references therein). In many of these instances, coupled positive and negative feedback loops, are hypothesized to yield hysteric relaxation oscillations.
Alternate hypothesis suggest that the essential ingredients of relaxation oscillators are a threshold device that enable a switch in direction, for example a bistable system, and a negative feedback loop. These ingredients, are for instance, the main components of the van der Pol oscillator in systems dynamics, and the Fitzhugh-Nagumo oscillator \cite{cite-key-Fitzhugh}, Morris-Lecar \cite{cite-key-lacer} oscillators studied in 
neurobiology and the comparator based extension of the Pearson-Anson oscillator \cite{Pearson_1921}. In the system we studied extensively and presented in this article analogues to each component exist with the active follower force serving as the source of power.  

We propose that the stable long-time relaxation oscillations in our system can be modeled in a simpler manner by studying the swirl-lip phenomenon as a two time-scale problem - with one time constant being much larger than the other. The challenge here would be to identify the minimal set of variables and parameters and is the subject of current work. Nonetheless, equations (1)-(7) together constitute a complex but complete set of time-stepper equations that in conjunction with continuation and bifurcation techniques may be used to analyze the stability of time dependent states as well as identify the critical points of onset.

\section{Summary}

In this paper we have used a geometrically nonlinear continuum rod model to analyze the stable two and three dimensional oscillations of strongly constrained pre-stressed rods activated by distributed follower forces. A dissipative component resisting rod motion was imposed by assuming that the ambient fluid medium exerts a resisting quadratic drag following Morison's form corresponding to flow fields characterized by $O(1)$ and higher Reynolds numbers. Nonetheless, based on our previous work on planar flapping with both linear and non-linear drag models, we expect our results  to change quantitatively but not qualitatively for a different drag force such as the linear drag valid at low Reynolds number.

Our results from this work can be summarized as follows. For rods buckled to bent and twisted base states, we observe a swirling motion around the end-to-end axis under any non-zero follower force. Moreover, a second characteristic time scale of this motion is identified by the rate in which swirling motion undergoes reversal of direction, or flipping. We analyzed the force-frequency behavior as a function of pre-stress, measured by end-to-end compression, as well as material elasticity. For the range of parameters examined here we identify a linear relationship between the force density and swirling frequencies. The flipping rate is found to be sensitive to the force density only when forces are small. When force intensity increases flapping becomes independent of the force and is only a function of compression and torsional deflection.    

The dynamical responses of animated slender rods that are pre-stressed  constitute a rich, involved tapestry. Tuning the pre-stress and rod elasticity allows us to choose stable solutions from possible planar (2D) and  twisted (3D) oscillating states. The connection to relaxation oscillations motivates future theoretical work 
that will help uncover the mechanistic and dynamical principles,  and allow for phenomenological physical extensions of the theory to understand the role of inertia and other forms of fluid drag on the eventual spatiotemporal patterns attained. Practically, our results suggest avenues by which pre-stress, elasticity and activity may be used to as knobs in exploiting active elasto-hydrodynamic instabilities to design synthetic macro-scale fluidic elements such as pumps or mixers.

\bibliography{mybibfile.bib}

\end{document}